\documentclass[8pt,biblatex]{article}
\usepackage[utf8]{inputenc}
\usepackage[hyphens,spaces,obeyspaces]{url}
\usepackage[colorlinks,allcolors=blue]{hyperref}
\usepackage[numbers]{natbib}
\usepackage{paralist}
\usepackage{amsmath,amsfonts}

\usepackage{booktabs}

\usepackage[math]{blindtext}
\usepackage{mwe}

\usepackage{multirow}
\usepackage{makecell}
\usepackage{acronym} 
\usepackage{paralist} 
\usepackage{colortbl}
\usepackage{tabto} 
\definecolor{LightCyan}{rgb}{0.90,0.90,0.90}

  \acrodef{ACID}    {Atomicity, Consistency, Isolation, Durability}
  \acrodef{ANSI}    {American National Standards Institute}
  \acrodef{API}     {Abstract Programming Interface}
  \acrodef{BASE}    {Basic Availability, Soft State, Eventual Consistency}
  \acrodef{BBO}     {Black-box Optimization}
  \acrodef{CART}    {Classification and Regression Tree}
  \acrodef{CPU}     {Central Processing Unit}
  \acrodef{CQL}     {Cassandra Query Language}
  \acrodef{DAS}     {Direct Attached Storage}
  \acrodef{DB}      {Database}
  \acrodef{DBA}     {Database Administrator}
  \acrodef{DBMS}    {Database Management System}
  \acrodef{DT}      {Decision Tree}
  \acrodef{FIFO}    {First In, First Out}
  \acrodef{GB}      {Gigabyte}
  \acrodef{GBDT}    {Gradient Boosting Decision Tree}
  \acrodef{HDD}     {Hard Disk Drive}
  \acrodef{HDFS}    {Hadoop Distributed File System}
  \acrodef{IP}      {Internet Protocol}
  \acrodef{ISO}     {International Organization for Standardization}
  \acrodef{IT}      {Information Technology}
  \acrodef{JSON}    {JavaScript Object Notation}
  \acrodef{JVM}     {Java Virtual Machine}
  \acrodef{KB}      {Kilobyte}
  \acrodef{MAE}     {Mean Absolute Error}
  \acrodef{MSE}     {Mean Squared Error}
  \acrodef{MB}      {Megabyte}
  \acrodef{ML}      {Machine Learning}
  \acrodef{MLlib}   {(Apache Spark) Machine Learning Library}
  \acrodef{MQL}     {MongoDB Query Language}
  \acrodef{ms}      {Milliseconds}
  \acrodef{NAS}     {Network Attached Storage}
  \acrodef{NoSQL}   {Non-relational SQL or Non-SQL}
  \acrodef{OLTP}    {Online Transaction Processing}
  \acrodef{op/s}    {operations per second}
  \acrodef{OS}      {Operating System}
  \acrodef{QoS}     {Quality of Service}
  \acrodef{RAM}     {Random Access Memory}
  \acrodef{RDBMS}   {Relational Database Management System}
  \acrodef{RF}      {Random Forest}
  \acrodef{RMSE}    {Root Mean Squared Error}
  \acrodef{SAN}     {Storage Area Network}
  \acrodef{SLA}     {Service-Level Agreement}
  \acrodef{SQL}     {Standard Query Language}
  \acrodef{SSD}     {Solid-State Drive}
  \acrodef{SSTable} {Sorted Strings Table}
  \acrodef{TD}      {Tuning Domain}
  \acrodef{XML}     {Extensible Markup Language}
  \acrodef{YAML}    {YAML Ain't Markup Language}
  \acrodef{YCSB}    {Yahoo! Cloud Serving Benchmark}

\title{NoSQL Database Tuning through Machine Learning}
\author{Florian Eppinger \\
University of Hagen \\ 
Hagen, Germany  \\
florian.eppinger@gmail.com \\
\and
 Uta Störl \\ University of Hagen \\
 Hagen, Germany \\
uta.stoerl@fernuni-hagen.de}

\date{} 

\begin{document}

\maketitle

\begin{abstract}
NoSQL databases have become an important component of many big data and real-time web applications. Their distributed nature and scalability make them an ideal data storage repository for a variety of use cases. While NoSQL databases are delivered with a default ''off-the-shelf'' configuration, they offer configuration settings to adjust a database's behavior and performance to a specific use case and environment. The abundance and oftentimes imperceptible inter-dependencies of configuration settings make it difficult to optimize and performance-tune a NoSQL system. There is no one-size-fits-all configuration and therefore the workload, the physical design, and available resources need to be taken into account when optimizing the configuration of a NoSQL database. This work explores Machine Learning as a means to automatically tune a NoSQL database for optimal performance. Using Random Forest and Gradient Boosting Decision Tree Machine Learning algorithms, multiple Machine Learning models were fitted with a training dataset that incorporates properties of the NoSQL physical configuration (replication and sharding). The best models were then employed as surrogate models to optimize the Database Management System's configuration settings for throughput and latency using a Black-box Optimization algorithm. Using an Apache Cassandra database, multiple experiments were carried out to demonstrate the feasibility of this approach, even across varying physical configurations. The tuned \ac{DBMS} configurations yielded throughput improvements of up to 4\%, read latency reductions of up to 43\%, and write latency reductions of up to 39\% when compared to the default configuration settings.
\end{abstract}


\section{Introduction}
\label{sec:introduction}
The rate at which data is created, used, and persisted increases rapidly. While \acp{RDBMS} continue to play an important role in today's technology environments, \ac{NoSQL} \acp{DB} have become an integral part of real-time analytics or big data applications \cite{Sadalage2012,Chen2019}. Choosing the best \ac{NoSQL} solution and developing the best physical design for a given use case can be a challenging task on its own \cite{Lourenco2015,Herrero2016,Qader2018}. Furthermore, \ac{NoSQL} technologies offer an abundance of configuration settings that allow the system administrator to adjust the \ac{DB} behavior to further meet the requirements of a particular use case. Many of the configuration parameters have an impact on the performance of the \ac{NoSQL} \ac{DB}, i.e., its throughput and latency.\par
\emph{Finding the configuration that maximizes throughput or minimizes latency for a given use case is complex, and \ac{DB} behavior is not always self-explanatory.} 
For example, one would assume that if a \ac{NoSQL} \ac{DB} is deployed on a significantly more powerful hardware configuration, its performance will increase accordingly. However, Preuveneers and Joosen have shown that this is not necessarily the case through research that demonstrates that moving a MongoDB \ac{DB} instance from low-end desktop hardware to high-end server hardware did not yield the expected results \cite{Preuveneers2020}.\par
Having the ability to predict performance measures for varying workloads and physical configurations, and to optimize \ac{DBMS} configuration settings accordingly, could be beneficial in various situations, e.g.:
\begin{compactitem}
\item Sudden spikes in user activity or processing needs may require quick action from \acp{DBA}. Intuition may lead to acquiring additional hardware resources, resulting in an oftentimes extensive implementation process. Having the ability to quickly evaluate the performance of various physical configurations, such as temporarily reducing the replication factor or increasing the node count, could significantly improve the ability to counteract the spike and comply with \acp{SLA}.
\item As an application matures, changes, or grows, the workload composition may also change. This, in turn, may require updates to the physical configuration or \ac{DBMS} configuration settings to achieve optimum performance. 
\item Hardware or software updates may require planned maintenance and outages of individual \ac{NoSQL} nodes. Having the ability to predict performance behavior, and the ability to quickly tune the new physical configuration for optimum performance, could be a valuable tool in such situations to maintain \acp{SLA} and ensure a good user experience.
\end{compactitem}
The configuration challenge is further amplified by \emph{a shortage of experienced resources}, recently highlighted in studies by Simplilearn\footnote{\href{https://www.simplilearn.com/why-nosql-skills-are-crucial-for-big-data-career-article}{https://www.simplilearn.com/why-nosql-skills-are-crucial-for-big-data-career-article} (visited on Oct.\ \(6^{th}\), 2022)} and Deloitte\footnote{\href{https://www2.deloitte.com/us/en/insights/industry/technology/data-analytics-skills-shortage.html}{https://www2.deloitte.com/us/en/insights/industry/technology/data-analytics-skills-shortage.html} (visited on Oct.\ \(6^{th}\), 2022)}.
Developing a schema and tweaking the \ac{DB} configuration is often handled by software developers, and is likely based on intuition or trial \& error rather than expert knowledge and hands-on experience. \par
Not only can an optimum configuration maximize performance, but it also \emph{reduces infrastructure and resource cost}. This, in turn, has a \emph{positive impact on sustainability}, which has become an essential aspect of business operations, corporate culture, and corporate identity.\par
The main objective of this study is to evaluate the ability of Machine Learning models to make performance predictions for varying \ac{DBMS} configurations \emph{and} DB physical configurations.  We thus make the following contributions:
\begin{compactitem}
    \item We analyze the quality of \ac{ML} models that are trained to predict performance metrics for varying workloads, physical configurations, and \ac{DBMS} configurations.
    \item We measure how the size of the training dataset influences the quality of these \ac{ML} models.
    \item We evaluate how features representing the physical configuration impact the quality of the \ac{ML} models.
    \item We explore the ability to tune the \ac{DBMS} configuration settings for a specific physical configuration using \ac{BBO} methods that utilize the fitted \ac{ML} models. 
\end{compactitem}

Section~\ref{sec:relatedwork} introduces related work discovered and reviewed during the literature review that was conducted in preparation for this study. Section~\ref{sec:methodology} briefly introduces the end-to-end methodology used by the authors and also defines the \ac{TD} in detail, which specifies the properties of the workload, the \ac{DBMS} configuration settings, as well as aspects of the physical \ac{DB} configuration that were considered within the scope of this study. Section~\ref{sec:trainingdata} introduces the training dataset and explains the approach that was used to generate it. Results and findings are presented and evaluated in section~\ref{sec:experimentalevaluation}. The document concludes with section~\ref{sec:discussion}, which discusses the results and makes suggestions for areas of future work.

\section{Related Work}
\label{sec:relatedwork}
A variety of proposals have been made to mitigate the performance tuning challenges outlined in section~\ref{sec:introduction} through automation.\par
Table \ref{tab:literaturereview:summary} summarizes and categorizes the related work that was reviewed as part of this study. Column \emph{Target} specifies the type of software that the tuning approach is focused on. Some of the methods are able to tune a variety of software solutions in a generic fashion \cite{Zhu2017,Wang2018}, while others are geared specifically toward \ac{RDBMS} or \ac{NoSQL} technology. Column \emph{Tuning Domain} defines what aspect of the system is considered during performance tuning. For \ac{DB} solutions, the work can be categorized into approaches that tune \ac{DB} performance via the \ac{DBMS} configuration settings \cite{Khattab2015,Zhang2019,Aken2021,Xiong2017} and solutions that tune \ac{DB} performance via the physical design of the \ac{DB} \cite{Cruz2013,Bermbach2015,Farias2016}, where the physical design represents aspects such as the schema design, replication, sharding, etc. Column \emph{State} specifies \emph{when} the tuning process takes place. \emph{Offline} indicates that the \ac{DB} is tuned in an offline state, i.e. the database is not monitored and configuration and design are not adjusted in real-time. Instead, the system is used to analyze a given configuration or make recommendations to a \ac{DBA} who then actually implements the changes. \emph{Online}, on the other hand, means that the database is monitored and configuration and design are adjusted autonomously while the system is active. \emph{Tuning Methods} include Control Theory, Expert Systems, and a variety of Machine Learning algorithms.

\begin{table}[hbt!]
\centering
\small{
\begin{tabular}{lllll}
\toprule
\textbf{} & \textbf{Target} & \textbf{Tuning Domain} & \textbf{State} & \textbf{Tuning Method} \\
\midrule
\makecell[l]{\emph{BestConfig} \\ \cite{Zhu2017}} & Generic & Configuration & Offline & \makecell[l]{Divide \& Diverge Sampling,\\Recursive Bond Search} \\
\midrule
\makecell[l]{\emph{SmartConfig} \\ \cite{Wang2018}} & Generic & Configuration & Online & Control Theory \\
\midrule
\makecell[l]{\emph{MAG} \\ \cite{Khattab2015}} & \ac{RDBMS} & \ac{DBMS} Config. & Offline & Neural Network\\
\midrule
\makecell[l]{\emph{CDBTune} \\ \cite{Zhang2019}} & \makecell[l]{\ac{RDBMS}\\(Cloud)} & \ac{DBMS} Config. & Online & \makecell[l]{Deep Reinforcement Learning}\\
\midrule
\makecell[l]{\emph{OtterTune} \\ \cite{Aken2021}} & \ac{RDBMS} & \ac{DBMS} Config. & \makecell[l]{Online,\\Offline} & \makecell[l]{Gaussian Process Regression,\\Deep Neural Networks, Deep\\Deterministic Policy Gradient}\\
\midrule
\makecell[l]{\emph{MET} \\ \cite{Cruz2013}} & \ac{NoSQL} & \ac{DB} Design & Online & Expert System\\
\midrule
\makecell[l]{\emph{Bermbach et al.} \\ \cite{Bermbach2015}} & \ac{NoSQL} & \makecell[l]{\ac{DB} Design\\(Schema)} & Offline & Expert System\\
\midrule
\makecell[l]{\emph{Farias et al.} \\ \cite{Farias2016}} & \ac{NoSQL} & \makecell[l]{\ac{DB} Design\\(Distribution)} & Offline & \makecell[l]{Linear Regression,\\Gradient Boosting Machine}\\
\midrule
\makecell[l]{\emph{ATH} \\ \cite{Xiong2017}} &\ac{NoSQL} & \ac{DBMS} Config. & Offline & \makecell[l]{Ensemble Algorithm and\\Genetic Algorithm}\\
\midrule
\makecell[l]{\emph{PaSTA} \\ \cite{Preuveneers2020}} & \ac{NoSQL} & \makecell[l]{\ac{DBMS} Config.,\\\ac{DB} Physical Design} & Online & Adaptive Hoeffding Trees\\
\midrule
\makecell[l]{\emph{ConfAdvisor} \\ \cite{Chen2021}} & \ac{NoSQL} & \makecell[l]{\ac{DBMS} Config.,\\\acs{OS} Kernel Config.} & Online & Black-box Optimization\\
\midrule
\rowcolor{LightCyan}
\emph{This Study} & \ac{NoSQL} & \makecell[l]{\ac{DBMS} Config.,\\\ac{DB} Physical Design} & Offline & \makecell[l]{Ensemble Algorithm and\\ Black-box Optimization}\\
\bottomrule
\end{tabular}}
    \caption{Summary of related work}
    \label{tab:literaturereview:summary}
\end{table}

One of the key differences between this work and related performance-tuning work for \ac{NoSQL} systems is that the \acl{TD} includes \emph{both} \ac{DBMS} configuration settings as well as the \ac{DB} physical design in form of sharding and replication. While Preuveneers and Joosen consider both aspects \cite{Preuveneers2020}, their technique differs in several ways. First, Preuveneers and Joosen utilize multiple predefined tactics in an attempt to tune the \ac{NoSQL} system \emph{online}. Second, the authors utilize the Adaptive Hoeffding Tree \ac{ML} model to map \ac{DB} scenarios consisting of workload metrics, resource utilization, physical configuration, etc. to an ideal \ac{DBMS} configuration and \ac{DB} physical design. This study, on the other hand, evaluates \ac{ML} techniques to fit ensemble models to predict \ac{DB} performance. These models are then used by \ac{BBO} algorithms to find an optimum \ac{DBMS} configuration for a given workload and physical configuration. Consequently, the methods used in this paper are related much closer to the approach presented by Xiong et al. \cite{Xiong2017}, except that this work utilizes Apache Cassandra instead of HBase, employs different optimization algorithms, and also includes features for the physical configuration.\par

\section{Methodology}
\label{sec:methodology}
Apache Cassandra\footnote{\href{https://cassandra.apache.org/}{https://cassandra.apache.org/} (visited on Dec.\ \(22^{th}\), 2022)} (``Cassandra'') was selected as the basis for experiments, and a prototypical implementation and domain knowledge was established, specifically focusing on the \ac{DBMS} configuration settings as well as aspects of the physical configuration. A tuning domain was defined using this domain knowledge, and a training dataset was generated using a sample database. This training dataset was then transformed into input for \ac{DT} \ac{ML} algorithms to fit various \ac{ML} models. The models' objective is to accurately predict the performance metrics of the database for a given workload, \ac{DBMS} configuration, and physical design. The performance of the \ac{ML} models was evaluated using commonly accepted quality measures. The predictions of the \ac{ML} models were then used to find optimized \ac{DBMS} configuration values for a given workload and physical configuration using a \acl{BBO} algorithm. Finally, actual performance gains of the optimized configurations were evaluated through additional experiments that applied the suggested \ac{DBMS} configuration and compared it against the performance of the default \ac{DBMS} configuration.\par
\begin{figure}[hbt!]
\centering{
\def\svgwidth{280bp}
\tiny{
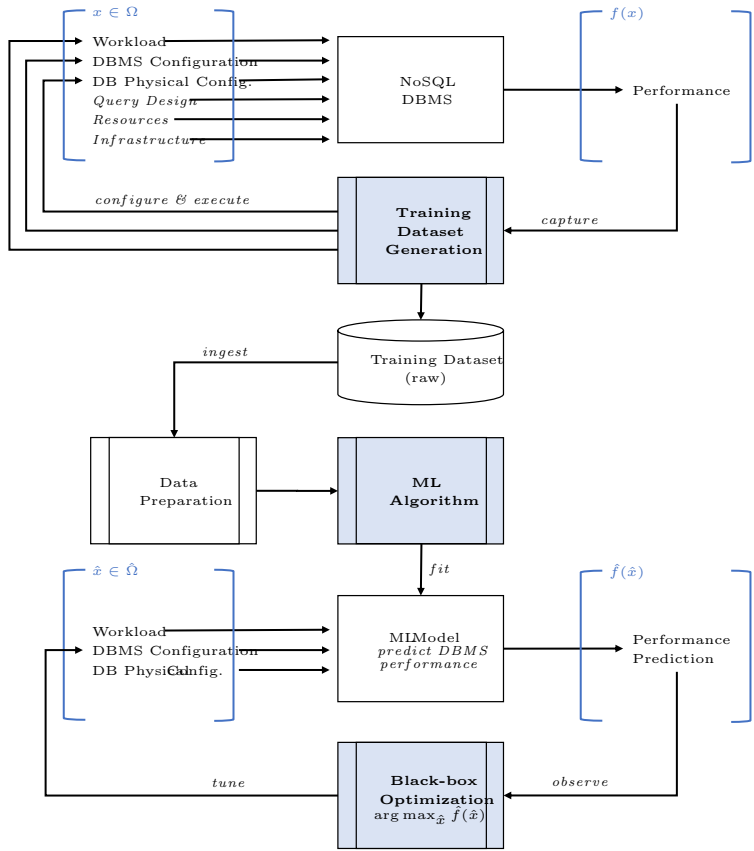
}
 \caption{Overview of the methodology}
 \label{fig:intro:methodology:overview}
}
\end{figure}

Figure~\ref{fig:intro:methodology:overview} provides an overview of the methodology and defines the terminology that is used throughout the remainder of this document:

\NumTabs{12} 
\tab \(x\) \tab := \tab A multi-dimensional input vector representing all factors that\\
\tab\tab\tab\tab influence the performance of the \ac{DB} (workload information, \ac{DBMS}\\
\tab\tab\tab\tab configuration, etc.).\\
\tab\tab \(\Omega\) \tab := \tab The domain of \(x\).\\
\tab\tab \(f(x)\) \tab :=\tab An unknown function that represents the actual \ac{DB} performance\\
\tab\tab\tab\tab for a given \(x \in \Omega\).\\
\tab\tab \(\hat{x}\) \tab := \tab A multi-dimensional input vector that contains values for some of\\
\tab\tab\tab\tab the factors that influence the performance of the database, specifically \\
\tab\tab\tab\tab workload information, \ac{DBMS}  configuration settings, and aspects of \\
\tab\tab\tab\tab the physical \ac{DB} design.\\
\tab\tab \(\hat{\Omega}\) \tab := \tab The domain of \(\hat{x}\), the \emph{tuning domain} (see section~\ref{sec:methodology:tuningdomain}).\\
\tab\tab \(\hat{f}(\hat{x})\) \tab := \tab An \ac{ML} model that approximates function f to determine the \ac{DB}\\
\tab\tab\tab\tab performance for a given \(\hat{x} \in \hat{\Omega}\).\\

\subsection{Tuning Domain}
\label{sec:methodology:tuningdomain}
As highlighted in section~\ref{sec:relatedwork}, the majority of previous work in the field of autonomous performance tuning for \ac{NoSQL} \acp{DBMS} is focused on either the \ac{DBMS} configuration \emph{or} the optimization of physical design aspects, such as the schema design or sharding strategy. Very little focus has been given to autonomous tuning of \ac{NoSQL} \acp{DB} using \acl{ML} with a feature set that includes both the \ac{DBMS} configuration \emph{and} aspects of the physical configuration. Both are important aspects when optimizing the performance of a database, and it may not always be apparent whether better performance can be achieved by a different physical configuration, a different \ac{DBMS} configuration, or a combination of both.\par
This study attempts to fit \ac{ML} models to make predictions for various \ac{DBMS} configuration settings and physical configurations. This section describes the \acl{TD}, i.e., the workload properties, the specific \ac{DBMS} configuration settings, as well as the aspects of the Cassandra physical configuration that were considered within the scope of this study. The \acl{TD} was then used to generate a training dataset for the \ac{ML} algorithms (see section~\ref{sec:trainingdata}).

Table \ref{table:sec:methodology:tuningdomain} provides a summary of the features that were considered in this study along with their respective tuning domains:

\begin{table}[ht]
\centering
\small{
\begin{tabular}{lllll}
  \toprule
\textbf{Feature} & \textbf{Feature Category} & \textbf{Domain}\\
  \midrule
  \texttt{wl\_read\_\%} & Workload & \((0\%-100\%)\) \\
  \texttt{wl\_write\_\%} & Workload & \((0\%-100\%)\) \\
  \texttt{trickle\_fsync} & \ac{DBMS} Configuration & \{true, false\}\\
  \texttt{key\_cache\_size\_in\_mb} & \ac{DBMS} Configuration & \(\{0, 2^0, ..., 2^5\}\)\\
  \texttt{row\_cache\_size\_in\_mb} & \ac{DBMS} Configuration &  \(\{0, 20, 40, 60, ..., 200\}\)\\
  \texttt{commitlog\_segment\_size\_in\_mb} & \ac{DBMS} Configuration & \(\{2^2, ..., 2^6\}\)\\
  \texttt{concurrent\_reads} & \ac{DBMS} Configuration & \(\{2^1*disks, ..., 2^5*disks\}\)\\
  \texttt{concurrent\_writes} & \ac{DBMS} Configuration & \(\{2^1, ..., 2^8\}\)\\
  \texttt{memtable\_heap\_space\_in\_mb} & \ac{DBMS} Configuration & \(\{2^{-5}*heap, ..., 2^{-1}*heap\}\)\\
  \texttt{node\_count (n)} & Physical Design & \(\{2, 3, 4\}\)\\
  \texttt{replication\_factor (rf)} & Physical Design & \(\{1, 2, 3, 4\}\)\\
  \bottomrule
\end{tabular}}
\caption{Features included in the Tuning Domain}
\label{table:sec:methodology:tuningdomain}
\end{table} 

\paragraph{Workload}
\label{sec:methodology:tuningdomainn:workload}
To fully dissect a workload and identify all aspects having an impact on performance is beyond the scope of this study. However, to follow the approach chosen in related work, such as \cite{Cruz2013} and \cite{Preuveneers2020}, and to gain insight into the effect of workload composition on performance, three workloads with different distributions of read and write operations were included:
\begin{compactitem}
\item 50\% read using the primary key, 50\% write (\emph{readwrite})
\item 95\% read using the primary key, 5\% write (\emph{readyheavy})
\item 5\% read using the primary key, 95\% write (\emph{writeheavy})
\end{compactitem}

\paragraph{\ac{DBMS} Configuration Settings}
\label{sec:methodology:tuningdomain:dbmsconfigurationsettings}
Apache Cassandra exposes a large amount of \ac{DBMS} configuration settings to the user. The main configuration file (\emph{cassandra.yaml}) alone contains far more than one hundred configuration settings. To reduce the complexity of this study, a subset of performance-relevant configuration settings was selected through research and targeted experiments.

\paragraph{Physical Design}
\label{sec:methodology:tuningdomain:dbphysicalconfig}
To account for aspects of the physical configuration, this study considers both \emph{sharding} and \emph{replication}. One of the fundamental concepts of distributed data stores is database \emph{sharding}. Sharding is a type of partitioning, i.e., a data set is broken into smaller subsets. While partitioning does not require that the dataset be broken out across multiple nodes in a clustered environment, this fragmentation is implied by sharding \cite{Sadalage2012}. To understand the ability of \ac{ML} models to make predictions based on the sharding configuration of a database, various sharding configurations were evaluated (node count \texttt{n} \(\in\) \{2,3,4\}). A single node was purposefully omitted under the assumption that it is not a common scenario. Database \emph{replication} within the context of this work is the practice of storing the same data in multiple locations. The replication factor (\texttt{rf}) determines on how many nodes each data record is stored.

\subsection{Tuning Subdomains}
\label{sec:methodology:tuningdomain:subdomains}
One of the items that sets this study apart from the related work presented in section~\ref{sec:relatedwork} is the inclusion of sharding and replication features. To analyze the impact of these physical configuration settings on the accuracy of the \ac{ML} predictions and the models' ability to tune the \ac{DBMS} configuration, Table \ref{table:sec:methodology:machinelearning:trainingcontext:trainingcontexts} further categorizes the \acl{TD} into \ac{TD}1-\ac{TD}4, which define subsets of the \acl{TD} that focus on specific aspects of the feature set.

\begin{table}[ht]
  \centering
  \small{
\begin{tabular}{lccc}
  \toprule
\textbf{\ac{TD}} & \textbf{Physical Config.} & \textbf{Workload} & \textbf{\ac{DBMS} Config.} \\
  \midrule
  \emph{\ac{TD}1} & * & * & * \\
  \midrule
  \emph{\ac{TD}2} & \emph{*} & \texttt{wl\_read\_\%}=fixed, \texttt{wl\_write\_\%}=fixed & \emph{*} \\
  \midrule
  \emph{\ac{TD}3} &  \texttt{n}=fixed, \texttt{rf}=fixed & \emph{*} & \emph{*} \\
  \midrule
  \emph{\ac{TD}4} & \texttt{n}=fixed, \texttt{rf}=fixed & \texttt{wl\_read\_\%}=fixed, \texttt{wl\_write\_\%}=fixed & \emph{*} \\
  \bottomrule
\end{tabular}}
\caption{Definition of the tuning subdomains} 
\label{table:sec:methodology:machinelearning:trainingcontext:trainingcontexts}
\end{table} 

\ac{TD}1 represents the entire tuning domain, i.e. all of the features introduced in Table \ref{table:sec:methodology:tuningdomain}. This \acl{TD} is the most complex as it attempts to fit \ac{ML} models that are able to make predictions for varying workloads, \ac{DBMS} configurations, and physical configurations. \ac{TD}2 reduces complexity by removing the workload features from the feature vector. It was designed to train \ac{ML} models that are able to make predictions for a fixed workload. \ac{TD}3, on the other hand, excludes the physical configuration features from the feature vector. \ac{TD}4 only considers the \ac{DBMS} configuration settings, omitting both the workload and physical configuration features.\par
\ac{TD}1-\ac{TD}4 are utilized in section~\ref{sec:experimentalevaluation:mlmodelquality} to evaluate how workload and physical configuration features impact the accuracy of the fitted \ac{ML} models.

\subsection{Tuning Methodology}
\label{sec:methodology:tuningmethodology}
The methodology applied in this study used \acf{ML}, more specifically \ac{RF} and \ac{GBDT} \ac{ML} algorithms to develop models that are able to make predictions about the performance of a \ac{DB}. Leveraging the \ac{ML} models' predictions, a \ac{BBO} algorithm is then utilized to search the optimum \ac{DBMS} configuration for a given workload and \ac{DB} physical configuration.\par

\paragraph{\acf{ML}}
\label{sec:methodology:tuningmethodology:ml}
Predicting performance metrics of the \ac{DB} is a form of regression analysis. Whether latencies or throughput, predictions represent real numbers. \emph{\ac{DT} learning} is a powerful supervised \ac{ML} methodology that can be used for both classification and regression and was chosen as the foundation for the \ac{ML} objective of this study due to the strengths and benefits outlined by Géron \cite{Geron2017}, including their effectiveness and unbiased nature. \aclp{DT} also require very little data preparation and feature scaling is typically not necessary, and they can be used with \emph{Ensemble Methods}. Rather than making a single model in hope of having found \emph{the best} \ac{ML} model for a particular learning objective, Ensemble Methods in \ac{ML} combine the prediction of multiple \ac{ML} models to make a final prediction. Two popular ensemble methods are \emph{Bootstrap Aggregation (Bagging)} and \emph{Boosting}. To evaluate both Bagging and Boosting methods, this study considered \acf{RF} and \acf{GBDT} \ac{ML} algorithms.\par
\emph{\acfp{RF}} are a popular ensemble \ac{ML} algorithm that combines multiple \aclp{DT} into a single predictive model. While individual \aclp{DT} are prone to overfitting when grown to a large depth, it has been shown using the Strong Law of Large Numbers that \acp{RF} converge with an increased number of trees in the ensemble \cite{Breiman2001}. \acp{RF} use Bagging to split the training dataset into subsets that are used to train the individual \ac{DT} models that are used to predict the outcome. In addition to the randomness introduced by Bagging, the \ac{RF} algorithm also randomly selects the features that are considered for a split at a given node to increase the variation among the individual \acp{DT} \cite{Breiman2001}.\par
The \acf{GBDT} algorithm uses a gradient-based Boosting methodology to develop an ensemble model. \emph{Gradient Boosting} iteratively improves the ensemble model by adding \ac{ML} models that are geared at minimizing the loss function using gradient descent \cite{Friedman2001, Natekin2013}. At each step in the sequence, the \ac{GBDT} constructs a new decision tree attempting to offset the overall ensemble loss by correlating the new tree to the negative gradient of its loss function. Unlike \acp{RF}, the likelihood of overfitting the ensemble model increases with the number of trees in the \ac{GBDT} ensemble \cite{Fafalios2020}.\par

\paragraph{\acf{BBO}}
\label{sec:methodology:tuningmethodology:bbo}
The objective of the \ac{ML} model is to accurately predict a \ac{DB} performance metric for a given workload, \ac{DBMS} configuration and physical design. In this sense, once fitted, it can be considered a \emph{surrogate model} for the actual performance behavior of the database. The model itself cannot find an optimum or near optimum configuration. Instead, the \ac{ML} model can be considered a black-box interface that can respond with a performance prediction for a given set of inputs. Because the model itself does not expose any meaningful information that could be used to find an optimum using a gradient-based optimization approach, this study explores \ac{BBO} as a means to find an optimum configuration over the surrogate \ac{ML} model.\par
A simple example of \ac{BBO} algorithms are \emph{Hill-Climbing algorithms}, or Local Optimization \cite{Johnson1989}. A hill-climbing algorithm is an iterative search that typically starts with a randomly or heuristically chosen value \(x^{opt} \in \Omega\). It then performs a local search, i.e. slightly adjust the original value to create \(x^{new} \in \Omega\). If \(f(x^{new})\) is deemed better than \(f(x^{opt})\), then \(x^{opt}\) is replaced with \(x^{new}\), otherwise a different \(x^{new}\) is chosen. This process is repeated until a certain number of steps have been executed, or a sufficiently good value \(f(x^{opt})\) has been found. The obvious drawback of the \emph{hill-climbing algorithms} is that it may be focused on a \emph{local optimum} rather than a \emph{global optimum}.\par
\emph{Simulated Annealing} is a popular \ac{BBO} algorithm that is modeled after the physical process of annealing, which means ``to subject (glass or metal) to a process of heating and slow cooling in order to toughen, reduce brittleness, or enhance adhesion'' \cite{anneal2022}. The goal of the Simulated Annealing algorithm is to find an ideal or near-ideal state in a system. The algorithm is very similar to the Local Optimization algorithm except that it attempts to avoid getting stuck in a non-global optimum by adding a random element to encourage further exploration outside of the current optimum \cite{Johnson1989}. The algorithm accomplishes this by allowing an occasional move in \emph{the wrong direction}, i.e., away from the current optimum. The likelihood that the algorithm allows this move decreases over time, just like it would in material science with a decreasing temperature. \par
Within the context of this study, the objective of the Simulated Annealing algorithm was to optimize the \ac{ML} model's input vector \(\hat{x} \in \hat{\Omega}\) for best performance \(\hat{f}(\hat{x})\) (maximum throughput or minimum latency).

\begin{table}[htb!]
\centering
\small{
\begin{tabular}{ccrrrrrrrrr}
\toprule
    \multicolumn{11}{c}{\textbf{Workload: readwrite (50\% read / 50\% write)}} \\
    \midrule
    \multirow{2}[0]{*}{\textbf{n}} & \multirow{2}[0]{*}{\textbf{rf}} & \multirow{2}[0]{*}{\textbf{Count}} & \multicolumn{2}{c}{\textbf{Throughput (ops/s)}} & \multicolumn{3}{c}{\textbf{Read Latency (ms)}} & \multicolumn{3}{c}{\textbf{Write Latency (ms)}} \\
          &     &       & \textbf{Max} & \textbf{Min} &       & \textbf{Min} & \textbf{Max} &       & \textbf{Min} & \textbf{Max} \\
    \midrule
    \textbf{4} & \textbf{4} &          3,068  &            77,670  &          30,962  &       & 10.1  & 34.8  &       & 3.6   & 8.4 \\
    \textbf{4} & \textbf{3} &          3,262  &            99,393  &          24,345  &       & 7.4   & 28.1  &       & 3.1   & 23.8 \\
    \textbf{4} & \textbf{2} &             960  &          101,398  &          44,708  &       & 6.8   & 20.8  &       & 4.2   & 8.3 \\
    \textbf{4} & \textbf{1} &          1,048  &          164,406  &          72,613  &       & 4.0   & 11.1  &       & 3.2   & 5.6 \\
    \textbf{3} & \textbf{3} &          1,444  &            83,787  &          30,152  &       & 8.3   & 34.2  &       & 5.3   & 9.7 \\
    \textbf{3} & \textbf{2} &          1,267  &            80,933  &          32,536  &       & 8.1   & 26.8  &       & 5.6   & 9.9 \\
    \textbf{3} & \textbf{1} &          1,575  &          158,199  &          66,520  &       & 4.0   & 12.4  &       & 3.0   & 6.6 \\
    \textbf{2} & \textbf{2} &          1,566  &            71,577  &          31,835  &       & 11.3  & 27.9  &       & 5.6   & 9.6 \\
    \midrule
    \multicolumn{11}{c}{\textbf{Workload: writeheavy (5\% read / 95\% write)}} \\
    \midrule
    \multirow{2}[0]{*}{\textbf{n}} &  \multirow{2}[0]{*}{\textbf{rf}} & \multirow{2}[0]{*}{\textbf{Count}} & \multicolumn{2}{c}{\textbf{Throughput (ops/s)}} & \multicolumn{3}{c}{\textbf{Read Latency (ms)}} & \multicolumn{3}{c}{\textbf{Write Latency (ms)}} \\
          &     &       & \textbf{Max} & \textbf{Min} &       & \textbf{Min} & \textbf{Max} &       & \textbf{Min} & \textbf{Max} \\
    \midrule
    \textbf{4} & \textbf{4} & 1,344  &            92,760  &          61,881  &       & 6.4   & 15.8  &       & 7.0   & 10.1 \\
    \textbf{4} & \textbf{3} & 1,433  &          111,018  &          67,962  &       & 4.9   & 11.3  &       & 5.9   & 8.5 \\
    \textbf{4} & \textbf{2} & 1,511  &          120,197  &          78,603  &       & 4.5   & 8.7   &       & 5.6   & 8.0 \\
    \textbf{4} & \textbf{1} & 1,221  &          153,518  &          93,332  &       & 4.5   & 10.0  &       & 4.2   & 8.3 \\
    \textbf{3} & \textbf{3} & 1,012  &          100,949  &          57,115  &       & 6.4   & 14.6  &       & 6.6   & 10.2 \\
    \textbf{3} & \textbf{2} & 1,251  &          110,267  &          69,294  &       & 4.8   & 8.7   &       & 5.9   & 8.4 \\
    \textbf{3} & \textbf{1} & 1,058  &          153,563  &          91,916  &       & 4.1   & 7.7   &       & 4.3   & 6.1 \\
    \textbf{2} & \textbf{2} & 984  &            83,530  &          51,386  &       & 7.1   & 11.3  &       & 7.8   & 11.0 \\
    \midrule
    \multicolumn{11}{c}{\textbf{Workload: readheavy (95\% read / 5\% write)}} \\
    \midrule
    \multirow{2}[0]{*}{\textbf{n}} &  \multirow{2}[0]{*}{\textbf{rf}} & \multirow{2}[0]{*}{\textbf{Count}} & \multicolumn{2}{c}{\textbf{Throughput (ops/s)}} & \multicolumn{3}{c}{\textbf{Read Latency (ms)}} & \multicolumn{3}{c}{\textbf{Write Latency (ms)}} \\
          &    &       & \textbf{Max} & \textbf{Min} &       & \textbf{Min} & \textbf{Max} &       & \textbf{Min} & \textbf{Max} \\
    \midrule
    \textbf{4} & \textbf{4} & 1,191  &            66,265  &          17,965  &       & 9.8   & 37.0  &       & 3.5   & 8.7 \\
    \textbf{4} & \textbf{3} & 1,301  &            95,018  &          27,835  &       & 7.2   & 22.8  &       & 3.0   & 7.6 \\
    \textbf{4} & \textbf{2} & 925  &            89,513  &          28,786  &       & 7.7   & 22.0  &       & 4.2   & 8.9 \\
    \textbf{4} & \textbf{1} & 1,120  &          168,778  &          45,281  &       & 4.3   & 12.4  &       & 3.0   & 8.0 \\
    \textbf{3} & \textbf{3} & 1,177  &            75,551  &          19,148  &       & 8.8   & 32.8  &       & 5.8   & 12.7 \\
    \textbf{3} & \textbf{2} & 1,056  &            67,266  &          20,682  &       & 9.8   & 29.5  &       & 6.8   & 13.1 \\
    \textbf{3} & \textbf{1} & 976  &          157,652  &          41,213  &       & 4.6   & 14.3  &       & 3.0   & 8.3 \\
    \textbf{2} & \textbf{2} & 1,007  &            49,955  &          14,224  &       & 12.6  & 44.2  &       & 8.6   & 25.5 \\
\bottomrule
    \end{tabular}}%
  \caption{Overview of the training dataset}
  \label{tab:trainingdataset:trainingdata:overview}%
\end{table}%

\section{Training Data}
\label{sec:trainingdata}
To train models with \ac{ML} algorithms, a dataset with training examples is required. A training example contains information about the workload, the values for \ac{DBMS} configuration settings, information about the physical configuration of the database, as well as information about the actual performance that was measured when the workload was executed against the \ac{DB} using this particular \ac{DBMS} configuration and physical configuration. \par
As illustrated in figure~\ref{fig:intro:methodology:overview}, the performance of a \ac{NoSQL} \ac{DB} depends on a variety of factors, including the hardware resources and the query design. The training dataset, which contains the actual performance measurements for a specific situation,  is sensitive to these factors and therefore cannot be directly utilized for a different hardware environment or \ac{NoSQL} technology. Furthermore, this study considers the number of nodes and the replication factor in the tuning domain and therefore differs from related work (e.g. \cite{Xiong2017}). Extending an existing training dataset by different physical configurations would require access to the original hardware and software environment. This was not an option, and therefore, a new dataset with the required properties was created.\par

\pagebreak

Each training example generated used the iterative five-step methodology:
\begin{compactenum}
\item The Cassandra process was stopped on all nodes before generating each training example. This was necessary because some configuration settings cannot be updated while the Cassandra process is running. It also allowed for the deletion of any data that may have been cached, both on-disk and in-memory, to ensure that subsequent training examples did not benefit from unintended caching mechanisms. 
\item The relevant \ac{DBMS} configuration settings and physical configuration were then updated. 
\item The Cassandra processes were restarted on all cluster nodes. 
\item Using a workload generator, a workload was executed against the Cassandra cluster,.
\item Relevant performance metrics were captured and stored along with metadata information that includes the values for current \ac{DBMS} configuration settings, the physical configuration, and properties of the workload. 
\end{compactenum}
The training dataset was created on the University of Hagen's \emph{Newton} cluster, which, at the time of this study, consisted of five individual server nodes, newton1-newton5, each with 16 processors (Intel®Xeon®CPU E5-2630 v3 @ 2.40GHz) and 32GB of RAM. Ubuntu Linux6 version 18.04.6 LTS (Bionic Beaver) served as the \ac{OS} on each server node. A custom process was used in combination with \emph{cassandra-stress}\footnote{\href{https://cassandra.apache.org/doc/latest/cassandra/tools/cassandra\_stress.html}{https://cassandra.apache.org/doc/latest/cassandra/tools/cassandra\_stress.html} (visited on Dec.\ \(22^{th}\), 2022)} to execute the workload on newton1. Apache Cassandra was installed on newton2 - newton5 to ensure that the workload process did not
compete for resources with the Cassandra processes.
A dataset consisting of 32,757 examples was generated, repeating the above five steps for each training example. Generating a single training example took approximately 55 seconds, totaling roughly 500 hours for the entire training dataset.\par
Table \ref{tab:trainingdataset:trainingdata:overview} aggregates the training examples by workload, the number of nodes (n) in the Apache Cassandra cluster, and the replication factor (rf). It lists key performance metrics captured for each group. Read and Write Latencies represent the minimum and maximum \emph{average} latencies for each group. The performance metrics of the training dataset varied significantly. As an example, the read latency of the readwrite workload on a four node cluster (n=4) with a replication factor of 3 (rf=3) ranged from 7.4 \ac{ms} to 28.1\ac{ms} depending on the chosen \ac{DBMS} configuration.\par

\section{Experimental Evaluation}
\label{sec:experimentalevaluation}
\subsection{\acs{ML} Model Quality}
\label{sec:experimentalevaluation:mlmodelquality}
The quality of \ac{ML} models depends on a variety of factors, including the size and quality of the dataset, feature engineering, as well as the \ac{ML} methodology, training algorithm and hyperparameters chosen to fit the model. In case of supervised \ac{ML} regression a variety of different quality measures exists, including the \ac{MAE} and  \ac{RMSE} \cite{Geron2017}.\par
Table \ref{tab:sec:results:mlmodelevaluation:modelefficiency:hyperparamters} lists the \ac{MAE} and \ac{RMSE} for \ac{DB} throughput and latencies. The measurements were established by training a total of 6 individual \ac{ML} models using \ac{TD}1, i.e. the entire training dataset as defined in section~\ref{sec:methodology:tuningdomain:subdomains}.

\begin{table}[ht]
  \centering
  \small{
    \begin{tabular}{lrrrrrr}
  \toprule
          & \multicolumn{2}{c}{\textbf{Overall Throughput}} & \multicolumn{2}{c}{\textbf{Read Latency}} & \multicolumn{2}{c}{\textbf{Write Latency}} \\
          & \textbf{\ac{RF}} & \textbf{\ac{GBDT}} & \textbf{\ac{RF}} & \textbf{\ac{GBDT}} & \textbf{\ac{RF}} & \textbf{\ac{GBDT}} \\
  \midrule
    \textbf{\ac{MAE}} & 2,810.940 & 2,880.110 & 0.307 & 0.279 & 0.203 & 0.189 \\
    \textbf{\ac{MAE} (\%)} & 3.290 & 3.370 & 2.940 & 2.730 & 3.030 & 2.890 \\
    \textbf{\ac{RMSE}} & 4,646.420 & 5,064.020 & 0.500 & 0.493 & 0.369 & 0.369 \\
  \bottomrule
    \end{tabular}}
  \caption{\ac{ML} Model Quality}
  \label{tab:sec:results:mlmodelevaluation:modelefficiency:hyperparamters}%
  \end{table}

The results highlight that the \ac{RF} and \ac{GBDT} algorithms produced models of similar quality. The \ac{GBDT} model outperformed the \ac{RF} algorithm when predicting read and write latencies. The \ac{RF} algorithm, on the other hand, yielded slightly better results when predicting the overall throughput performance measure. Both \ac{ML} algorithms were better at fitting models that predict latencies than they were at fitting models that predict overall throughput. This can intuitively be explained by the fact that optimizing overall throughput is an exercise that involves optimizing both read and write latencies and is, therefore, more complex. \par
It should be noted that the hyperparameters of the \ac{ML} models referenced in this study were tuned for each \ac{ML} model individually. The values for each hyperparameter varied based on the features, the tuning domain, the training dataset, and the performance metric (label) considered during a particular experiment. Hyperparameter tuning of each model had a significant impact on the accuracy of each model. For example, experiments conducted as part of this study showed that hyperparameter tuning an \ac{RF} model improved the \ac{MAE} measure by more than 300\% compared to a model that was trained with default hyperparameter values.\par
\begin{figure}[ht]
\centering{
\large{
\resizebox{1\textwidth}{!}{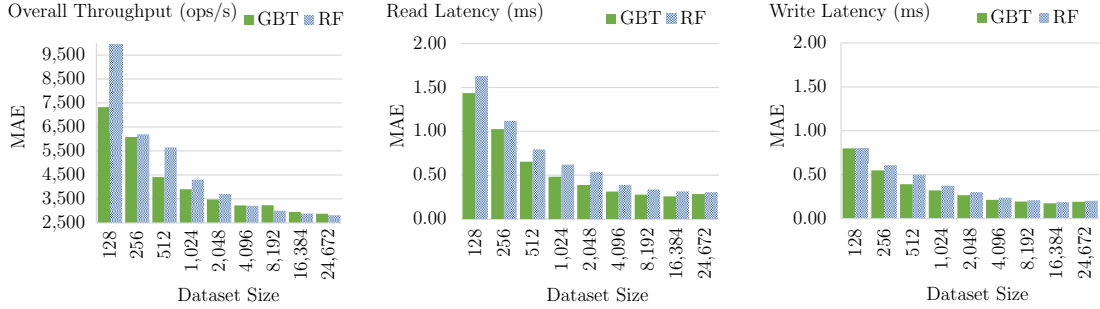}}
 \caption{Influence of the dataset size on the prediction accuracy of the \ac{ML} models}
 \label{fig:sec:results:mlmodelevaluation:modelefficiency:error_tds_size}
}
\end{figure}

The \ac{ML} model quality results outlined in Table \ref{tab:sec:results:mlmodelevaluation:modelefficiency:hyperparamters} were created with the entire dataset of 32,757 examples. Using a random split methodology, 75\% of the dataset was used to fit the models with holdout validation (see \cite{Geron2017}), and the remaining 25\% were used for testing. To understand the correlation between dataset size and \ac{ML} model prediction accuracy, various experiments were conducted with artificially reduced datasets. A total of 9 \ac{ML} models were trained for each of the performance measures using the following dataset sizes: 128, 256, 512, 1,024, 2,048, 4,096, 8,192, 16,384, 24,672. To ensure a fair evaluation and reduce the chance of randomly selecting a non-representative test distribution, the test dataset was kept at a fixed size of 8,000 examples. The results are shown in figure~\ref{fig:sec:results:mlmodelevaluation:modelefficiency:error_tds_size}. A training dataset size of 128 examples yielded a \ac{GBDT} model that predicted overall throughput with an \ac{MAE} of 7,326 and a \ac{RF} model that predicted overall throughput with an \ac{MAE} of 9,973. Increasing the training dataset size to just 1,024 records significantly reduced the \ac{MAE} values to 3,903 (\ac{GBDT}) and 4,305 (\ac{RF}). Additional accuracy could be gained by further increasing the dataset size. However, only minor improvements could be seen for read and write latencies with datasets exceeding 4,096 examples.

To evaluate how the TD1 \ac{ML} model quality  compares to the less complex \aclp{TD}, \ac{ML} models were trained for \ac{TD}2-\ac{TD}4 to predict write latencies. Four individual models were fitted for each combination of \ac{TD} and \ac{ML} algorithm using datasets with 128, 256, 512, and 1,024 training examples. For this experiment a test dataset size of 250 was used to determine the \ac{MAE} values for each of the models.\par
\begin{figure}[ht]
\centering{
\def\svgwidth{270bp}
\scriptsize{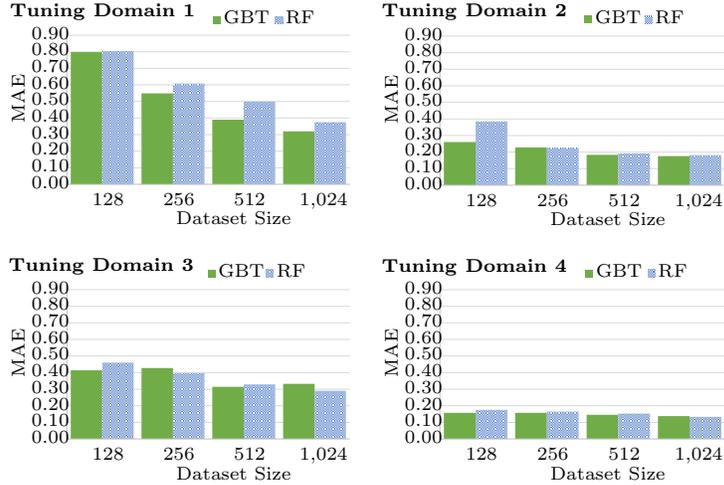}
 \caption{Influence of the dataset size on the write latency prediction accuracy of the \ac{ML} models trained with different \acsp{TD}}
 \label{fig:sec:results:mlmodelevaluation:modelefficiency:error_tds_size_tcs}
}
\end{figure}

The results are shown in figure~\ref{fig:sec:results:mlmodelevaluation:modelefficiency:error_tds_size_tcs} and confirm that the \ac{ML} models trained with \ac{TD}1 performed worse compared to \ac{TD}2-\ac{TD}4 when identically sized training datasets were used. A \ac{GBDT} model trained with 128 randomly selected examples from the \ac{TD}1 dataset yielded an \ac{MAE} of 1.42, which represents an error percentage of 13.62\% compared to 0.65 (5.88\%) for \ac{TD}2. The \ac{MAE} of the \ac{GBDT} model dropped to 0.39 (3.55\%) with 512 training examples for \ac{TD}2, while it took four times the number of training examples to reach comparable accuracy within \ac{TD}1. It is also worth noting that with 1,024 training examples, the \ac{TD}1 and \ac{TD}3 models were of similar quality despite the added complexity of the \ac{TD}1 model that is able to make predictions for eight different physical configurations. 

\subsection{Tuning performance with Black-box Optimization}
\label{sec:experimentalevaluation:bbo}
\begin{figure}[ht]
\centering{
\def\svgwidth{220bp}
\scriptsize{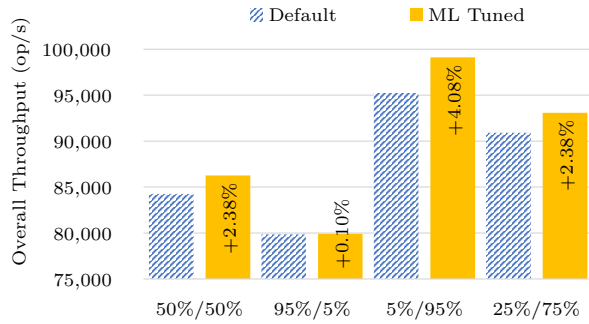}
 \caption{Actual throughput measurements for default and \acs{ML}/\acs{BBO}-tuned \acs{DBMS} configurations}
 \label{fig:results:bbooverthesurrogate:optimization_workload_throughput}
}
\end{figure}

\begin{figure}[ht]
\centering{
\def\svgwidth{350bp}
\scriptsize{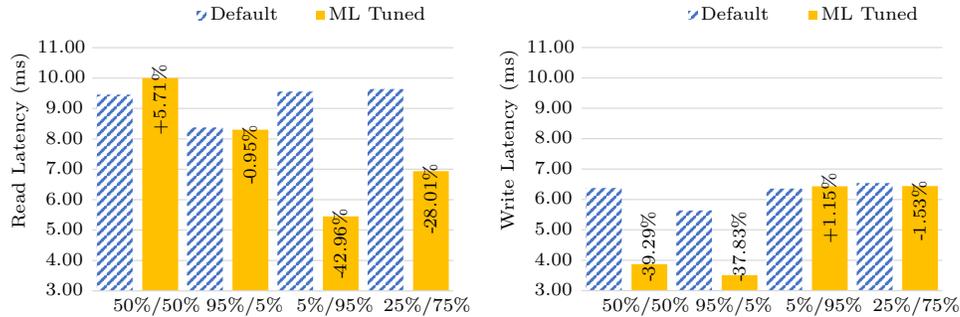}
 \caption{Actual latency measurements for default and \acs{ML}/\acs{BBO}-tuned \acs{DBMS} configurations}
 \label{fig:results:bbooverthesurrogate:optimization_workload_latencies}
}
\end{figure}

A search over the best \ac{TD}1 \ac{ML} model for the optimum \ac{DBMS} configuration was conducted using the Simulated Annealing \ac{BBO} algorithm. The search was carried out for various workloads and a physical configuration of four nodes, and a replication factor of three. The \ac{DBMS} configuration recommended by the optimization algorithm was then applied to the Cassandra cluster, respective workloads were executed, and results were captured. In addition to the \emph{readwrite} (50\% read, 50\% write), \emph{readheavy} (95\% read, 5\% write) and \emph{writeheavy} (5\% read, 95\% write) workloads, experiments were conducted with a workload that differed from the workloads used to train the \ac{ML} model (25\% read, 75\% write). For each performance measure, five independent experiments were carried out for each combination of \ac{DBMS} configuration and workload type, and average performance measure values were calculated. The results for all three performance measures are shown in figures \ref{fig:results:bbooverthesurrogate:optimization_workload_throughput} and \ref{fig:results:bbooverthesurrogate:optimization_workload_latencies}.

Overall throughput could be improved for all workloads, most notably for the write-heavy workload (4.08\%), while only a minor improvement was achieved for the read-heavy workload (0.10\%). The algorithm also found a configuration that increased overall throughput for the previously unseen 25\%/75\% workload. More drastic improvements were achieved for the read and write latencies. Tuned configurations could reduce the read latency by more than 42\% for write-heavy workloads and the write latency by more than 39\% for workloads with a significant amount of read operations. It should be noted that configurations that improved read latencies resulted in higher write latencies and vice versa, reducing overall throughput. However, these results clearly demonstrates that the \ac{ML} model was able to successfully derive knowledge which \ac{DBMS} configuration settings impact the performance measures and to what extent.\par
Next, we applied the optimization method to different physical configurations. Similar to the previous section, a search over the best \ac{TD}1 \ac{ML} model was conducted using the Simulated Annealing algorithm. However, this time the objective of the algorithm was to tune the performance by adjusting the \ac{DBMS} configuration under varying physical configurations. The measurements also included a physical configuration with two nodes (\texttt{n}=2) and a replication factor of one (\texttt{rf}=1) to analyze how well the methodology works for previously unseen physical configurations. While the training set included examples with two nodes and separate examples with a replication factor of one, it did not account for any examples with that particular combination. \par
\begin{figure}[ht]
\centering{
\def\svgwidth{280bp}
\scriptsize{
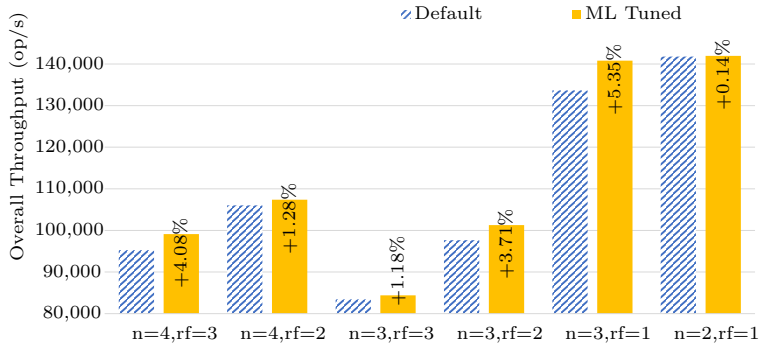
}
 \caption{Actual throughput for default and \acs{ML}/\acs{BBO}-tuned \acs{DBMS} configurations for various physical designs}
 \label{fig:results:bbooverthesurrogate:optimization_physicalconfig}
}
\end{figure}
\begin{figure}[ht]
\centering{
\def\svgwidth{280bp}
\scriptsize{
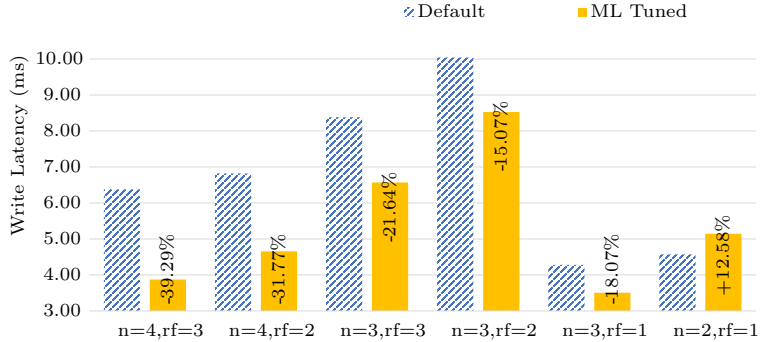
}
 \caption{Write latencies for default and \acs{ML}/\acs{BBO}-tuned \acs{DBMS} configurations under varying physical configurations}
 \label{fig:results:bbooverthesurrogate:optimization_physicalconfig_wl}
}
\end{figure}

Figure~\ref{fig:results:bbooverthesurrogate:optimization_physicalconfig} shows how the throughput performance of the \ac{ML}-tuned configuration compared with the default configuration (\emph{writeheavy} workload). Similarly, figure~\ref{fig:results:bbooverthesurrogate:optimization_physicalconfig_wl} shows how the write latency performance of the \ac{ML}-tuned configuration compared to the default configuration (\emph{readwrite} workload). It should be noted that the tuned \ac{DBMS} configuration settings varied across different physical configurations.

These results illustrate that the \ac{ML}-based tuning method could successfully tune the \ac{DB} for throughput and latency under a variety of physical configurations. However, it also highlights that it failed to optimize performance for physical configurations that were not previously encountered during the training phase of the \ac{ML} model. Rather than improving write latency, it increased by 12.58\% from $4.57\ac{ms}$ to $5.14\ac{ms}$. One reason for this may be that the \ac{ML} algorithms failed to extract and derive a meaningful trend that would allow accurate predictions for physical configurations that were not included in the training dataset. Another reason could be that the corresponding features were not encoded to cultivate this kind of predictive quality in the model. Instead of using the simple ordinal encoding scheme applied as part of this study, one may consider an ordinal encoding in combination with normalization over an extended range.\par
\begin{figure}[ht]
\centering{
\def\svgwidth{150bp}
\scriptsize{
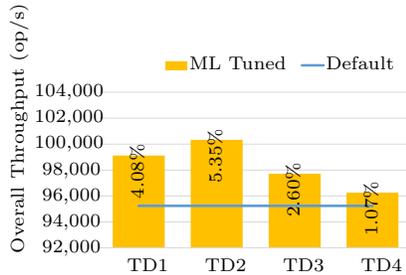
}
 \caption{Actual throughput measurements for \ac{DBMS} configurations that were optimized with \acs{ML} models trained with \acs{TD}1-\acs{TD}4}
 \label{fig:results:bbooverthesurrogate:optimization_tcs}
}
\end{figure}

We also analyzed models that were trained with the tuning subdomains \ac{TD}2-\ac{TD}4. Section~\ref{sec:experimentalevaluation:mlmodelquality} showed significant accuracy differences depending on what tuning domain was used to fit the \ac{ML} models, even when trained with datasets consisting of 1,048 examples. To evaluate the ability to tune the \ac{DBMS} configurations with these models, the Simulated Annealing algorithm was used to search for an optimum throughput configuration for each of the \acp{TD} using the \ac{ML} models fitted with 1,024 training examples. These configurations were then applied to the Cassandra cluster, the workload was executed, and performance metrics were captured. This process was repeated three times for each \ac{TD}, and average throughput results were calculated. The experiments were based on the \emph{writeheavy} workload (5\% read/95\% write), and results are shown in figure~\ref{fig:results:bbooverthesurrogate:optimization_tcs}. 

Based on the observations made in section~\ref{sec:experimentalevaluation:mlmodelquality} and because \ac{TD}4 clearly has the least complex feature vector, the expectation was that \ac{TD}2 and \ac{TD}4 would produce the most performant configurations and \ac{TD}1 would produce the least performant configuration. However, this was not the case. Instead, \ac{TD}1 outperformed both \ac{TD}3 and \ac{TD}4.\par
\begin{figure}[ht]
\centering{
\def\svgwidth{260bp}
\scriptsize{
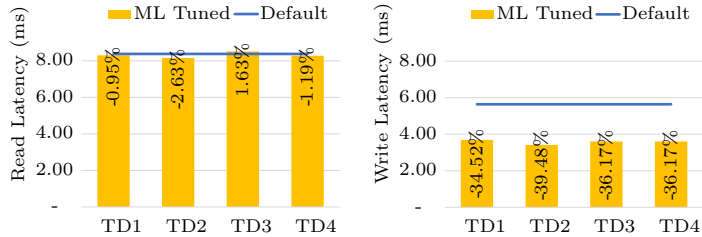
}
 \caption{Actual latency measurements for \ac{DBMS} configurations that were optimized with \acs{ML} models trained with \acs{TD}1-\acs{TD}4}
 \label{fig:results:bbooverthesurrogate:optimization_tcs_latency}
}
\end{figure}

The experiment was repeated using the \emph{readheavy} workload (95\% read/5\% write), this time optimizing latencies instead of throughput. The results are shown in figure~\ref{fig:results:bbooverthesurrogate:optimization_tcs_latency}.

These results aligned closer with expectations (\ac{TD}2 and \ac{TD}4 produce the best configurations). As illustrated in figure~\ref{fig:sec:results:mlmodelevaluation:modelefficiency:error_tds_size_tcs}, predictions of \ac{TD}2 and \ac{TD}4 models were roughly twice as accurate as predictions for \ac{TD}1 and \ac{TD}3 models, and thus additional research was conducted to analyze why the \ac{ML} models with higher accuracy were not able to produce \ac{DBMS} configurations that resulted in significantly better performance. It was ultimately determined that the ability to tune the \ac{DBMS} configuration depended equally on the \ac{ML} model's accuracy as it did on the quality of the training data. The 1,024 training examples were chosen randomly from the training dataset. The training dataset contained a significant number of examples that resulted in performance that was worse than the default configuration. While the \ac{ML} models for the less complex \acp{TD} were able to make more accurate predictions, they were not necessarily able to help the \ac{BBO} algorithm find an optimum configuration if they were not fitted with any of the high-performance training examples.

\section{Discussion}
\label{sec:discussion}
This study evaluated \acl{ML} and \acl{BBO} as a means to performance-tune \ac{NoSQL} \acp{DB}. The methodology involved fitting various \ac{RF} and \ac{GBDT} models using a custom training dataset to make predictions about the performance of a Cassandra \ac{NoSQL} \ac{DB}. When trained with the entire dataset, the \ac{ML} models yielded an \ac{MAE} of 3.29\% for throughput and 2.73\% and 2.89\% for read and write latencies, respectively. Omitting the workload from the feature set (\ac{TD}2) significantly improved accuracy, while omitting the features representing the physical configuration (\ac{TD}3) resulted in moderate improvements only. This observation may lead to the conclusion that \emph{the physical configuration adds less complexity to the model than varying workloads}. In fact, when using training datasets of 1,024 examples, the \ac{MAE} of the \ac{TD}1 and \ac{TD}3 \ac{GBDT} models were almost identical when predicting read latencies. Therefore, one may also conclude from these results that it is more efficient to train a single predictive model for multiple physical configurations than it is to train an individual model for each physical configuration.\par
The most accurate models were then used to optimize \ac{DBMS} configuration settings for a given workload and physical configuration using the Simulated Annealing optimization algorithm. The algorithm was able to find configurations for improved performance in almost all situations. For a physical design with four \ac{NoSQL} nodes and a replication factor of three, throughput improvements ranged from 0.10\% to 4.08\%, and latencies could be reduced by up to 42.96\% (read) and 39.29\% (write) depending on the workload composition. Similar results were achieved for varying physical configurations. Experiments were also conducted involving workloads with read/write compositions that differed from the training examples. The tuned configurations improved throughput and latency performance even for these previously unseen workloads.\par
While the tuned \ac{DBMS} configurations did result in performance improvements, none of the results matched or exceeded the performance of those training examples that exhibited the best performance. As an example, the \ac{BBO}-tuned \ac{DBMS} configuration (n=4, rf=3, writeheavy) resulted in maximum throughput of 103,965 \ac{op/s} compared to 95,240 \ac{op/s} for the default configuration. The most performant training example, however, resulted in 111,018 \ac{op/s}, implying an additional tuning potential of 6.5\%. Some of this can likely be attributed to the fluctuations of up to 8\% that were observed when measuring performance with cassandra-stress. Another potential explanation for this is the generalization capability of the \ac{ML} model. While generalization helps the \ac{ML} model to make more accurate predictions for previously unseen \ac{DBMS} configurations, it also regulates outlier configurations. These outliers represent both subpar but also optimum configurations.\par
Several discoveries and choices were made regarding technology and methodology. Furthermore, several areas remained unexplored due to time and resource constraints. The following list reflects some of these items and highlights potential areas for future work:
\begin{compactitem}
\item The \ac{BBO} optimization methodology implemented in this study targeted individual \ac{DB} performance measures (throughput \emph{or} read latency \emph{or} write latency). Performance objectives vary, and this may not always be desirable as performance tuning oftentimes involves finding a performance state that involves meeting multiple performance goals, e.g. reducing read latency \emph{and} guaranteeing a certain throughput. This could be accomplished by considering multiple performance measures when optimizing the \ac{DBMS} configuration. Xiong et al.\ approach this by combining multiple weighted performance measures into a single optimization objective \cite{Xiong2017}, an attempt that could be explored further.
\item It was also noted that the training dataset captures performance results that are specific to the hardware environment and technologies, i.e. the \ac{NoSQL} database, \ac{OS}, etc., that were used to generate it. This implies that the trained models are not able to make predictions for a different hardware environment or \ac{NoSQL} technology. Because the generation of training examples is such a time-consuming task, it appears worthwhile to explore the feasibility of transforming or scaling training examples or derived knowledge for a different hardware environment or technology stack.
\item The tuning scope of this study is limited. For example, 
the physical configuration is limited to the number of nodes and the replication factor. Many more aspects exist, including secondary indexes, schema design, various consistency levels, etc., that could be evaluated in more detail.
\item Another observation that could be made is that the attempt to tune the \ac{DBMS} configuration for previously unseen physical designs did result in performance that under-performed the default \ac{DBMS} configuration. The corresponding features were encoded as ordinal values. Changing the encoding scheme may improve these results.
\item It should also be noted that the tuning methodology utilized in this study treats the \ac{DBMS} configuration settings as global settings, i.e., the same configuration settings were used for all nodes of the Cassandra node ring. However, many settings can be configured individually for each cluster node. Research in this area presented by Cruz et al.\ \cite{Cruz2013} could be evaluated as an extension to the methodology outlined in this document.
\item As far as the \ac{ML} approach is concerned, this study evaluated \ac{RF} and \ac{GBDT} to develop a surrogate model for performance predictions. Various other \ac{ML} algorithms exist, and others are likely that would exhibit a similar or better prediction quality. 
 Similarly, various other \ac{BBO} algorithms exist, some of which have been shown to produce better results than the Simulated Annealing algorithms \cite{Chen2021}. A more systematic evaluation of different \ac{BBO} algorithms could be carried out.
\end{compactitem}

\bibliographystyle{alpha}
\bibliography{bibliography}

\end{document}